\documentclass[%
reprint,
superscriptaddress,
 amsmath,amssymb,
 aps,
]{revtex4-1}

\usepackage{graphicx}% Include figure files
\usepackage{dcolumn}% Align table columns on decimal point
\usepackage{bm}% bold math
\usepackage{xcolor}
\usepackage[normalem]{ulem}

\newcommand{\new}[1]{\textcolor{black}{{}#1}}

\usepackage[pagebackref=false]{hyperref}

\begin{document}

\author{L.~Banszerus}
\email{luca.banszerus@rwth-aachen.de.}
\affiliation{JARA-FIT and 2nd Institute of Physics, RWTH Aachen University, 52074 Aachen, Germany,~EU}%
\affiliation{Peter Gr\"unberg Institute  (PGI-9), Forschungszentrum J\"ulich, 52425 J\"ulich,~Germany,~EU}

\author{A.~Rothstein}
\affiliation{JARA-FIT and 2nd Institute of Physics, RWTH Aachen University, 52074 Aachen, Germany,~EU}%

\author{T.~Fabian}
\affiliation{Institute for Theoretical Physics, TU Wien, 1040 Vienna, Austria, EU}

\author{S.~M\"oller}
\author{E.~Icking}
\affiliation{JARA-FIT and 2nd Institute of Physics, RWTH Aachen University, 52074 Aachen, Germany,~EU}%
\affiliation{Peter Gr\"unberg Institute  (PGI-9), Forschungszentrum J\"ulich, 52425 J\"ulich,~Germany,~EU}

\author{S.~Trellenkamp}
\author{F.~Lentz}
\affiliation{Helmholtz Nano Facility, Forschungszentrum J\"ulich, 52425 J\"ulich,~Germany,~EU}

\author{D.~Neumaier}
\affiliation{AMO GmbH, Gesellschaft f\"ur Angewandte Mikro- und Optoelektronik, 52074 Aachen, Germany, EU}
\affiliation{University of Wuppertal, 42285 Wuppertal, Germany, EU}

\author{K.~Watanabe}
\affiliation{Research Center for Functional Materials, 
National Institute for Materials Science, 1-1 Namiki, Tsukuba 305-0044, Japan}
\author{T.~Taniguchi}
\affiliation{International Center for Materials Nanoarchitectonics, 
National Institute for Materials Science,  1-1 Namiki, Tsukuba 305-0044, Japan}%
\author{F.~Libisch}
\affiliation{Institute for Theoretical Physics, TU Wien, 1040 Vienna, Austria, EU}
\author{C.~Volk}
\author{C.~Stampfer}
\affiliation{JARA-FIT and 2nd Institute of Physics, RWTH Aachen University, 52074 Aachen, Germany,~EU}%
\affiliation{Peter Gr\"unberg Institute  (PGI-9), Forschungszentrum J\"ulich, 52425 J\"ulich,~Germany,~EU}%

\title{Electron-hole crossover in gate-controlled bilayer graphene quantum dots}

\date{\today}% It is always \today, today,


\begin{abstract} 
%% NEW CS
Electron and hole Bloch states in gapped bilayer graphene exhibit topological orbital magnetic moments with opposite signs near the band edges, which allows for tunable valley-polarization in an out-of-plane magnetic field. This intrinsic property makes electron and hole quantum dots (QDs) in bilayer graphene interesting for valley and spin-valley qubits. Here we show measurements of the electron-hole crossover in a bilayer graphene QD, demonstrating the opposite sign of the orbital magnetic moments associated with the Berry curvature.
%opposite sign of the topological orbital magnetic moments. 
Using three layers of metallic top gates, we independently control the tunneling barriers of the QD while tuning the occupation from the few-hole regime to the few-electron regime, crossing the displacement-field controlled band gap. The band gap is around 25 meV, while the charging energies of the electron and hole dots are between 3-5~meV. The extracted valley $g$-factor is around 17 and leads to opposite valley polarization for electron and hole states at moderate $B$-fields. Our measurements 
agree well with tight-binding calculations for our device.

\end{abstract}

\maketitle

Carbon-based quantum dots (QDs) are attracting attention as promising hosts for spin, valley or spin-valley qubits. While spin qubits - the central building blocks of spin-based quantum computing~\cite{Loss1998Jan} - have already been intensively studied in III/V heterostructures~\cite{Petta2005Sep,Nowack2011Sep} and silicon~\cite{Yoneda2017Dec,Watson2018Feb,Zajac2018Jan}, the valley degree of freedom (where present) has been considered mainly as an undesirable constraint. %disturbing.
This is beginning to change and  proposals of valley-based quantum computation~\cite{Rycerz2007Feb,Culcer2012Mar,Rohling2012Aug} have led to the demonstration of spin-valley qubits in carbon nanotubes~\cite{Laird2013Jul} and more recently in silicon~\cite{Penthorn2019Nov}.
Bilayer graphene (BLG) could play an important role in this development.
Besides the small spin-orbit interaction and the weak hyperfine coupling~\cite{Trauzettel2007Feb}, the  tunability of the valley dependent topological orbital magnetic moment in gapped BLG~\cite{Lee2020Mar} offers new knobs for controlling the valley degree of freedom.
%ling spin-valley qubits. 
As the topological orbital magnetic moment is associated with the Berry curvature, it has different signs for the different valleys, as well as for electrons and holes and causes a large and adjustable effective valley $g$-factor~\cite{Knothe2020Jun,Lee2020Mar}.
In contrast to silicon, where the lifting of the valley degeneracy is set by the confining potential, the large valley $g$-factor in BLG offers control of the valley splitting and allows to achieve full valley polarization as a function of magnetic field and quasi-particle index (electrons or holes).
This, together with the tunable band gap, 
makes BLG interesting for the implementation of spin-valley qubits based on electron and hole QDs.

Quantum dots in single-layer and bilayer graphene have been investigated intensively in the last decade. First QD devices were carved out from 
individual flakes~\cite{Ihn2010Mar} and allowed to study excited states spectra~\cite{Schnez2009,Volk2011Aug}, spin-states~\cite{Guttinger2010Sep} and charge relaxation~\cite{Volk2013Apr}. 
However, these devices were heavily affected by edge disorder~\cite{Bischoff2012Nov, Engels2013Aug}, which limited the control over the QD occupation, making  it difficult to identify the electron-hole crossover~\cite{Guttinger2009Jul}.
Recent technological improvements that combine the encapsulation of BLG in hexagonal boron nitride (hBN)~\cite{Engels2014Sep} with a graphite back gate~\cite{Overweg2018Jan} have opened the door to gate-controlled quantum point contacts~\cite{Kraft2018Dec,Banszerus2020May} and QDs~\cite{Eich2018Jul,Banszerus2018Aug}. Excited state spectroscopy~\cite{Kurzmann2019Jul}, charge detection~\cite{Kurzmann2019Aug} and single-electron occupation of single~\cite{Eich2018Jul} and double QDs~\cite{Banszerus2020Mar} have been shown for such BLG quantum devices.
Common to all these devices is that a set of gates (the so-called split gates) is used to form either a p-doped or n-doped channel in which an additional finger gate forms a p-n-p or n-p-n band profile.
The two naturally occurring pn-junctions serve as tunneling barriers of the gated-defined QD. This technology, however, limits the control of the tunneling rates and requires the QD to have the opposite polarity as the neighboring reservoirs~\cite{Eich2018Jul,Eich2018Jul,Banszerus2018Aug,Kurzmann2019Aug,Kurzmann2019Jul,Banszerus2020Mar,Banszerus2020Jun}.
The lack of independent control of the dot occupation explains why it has remained a challenge to demonstrate the direct transition from the few-electron to the few-hole regime (as in carbon nanotube QDs~\cite{Jarillo-Herrero2004May,Kuemmeth2008Mar}) in BLG QD devices.

\begin{figure*}[!thb]
\centering
\includegraphics[draft=false,keepaspectratio=true,clip,width=1\linewidth]{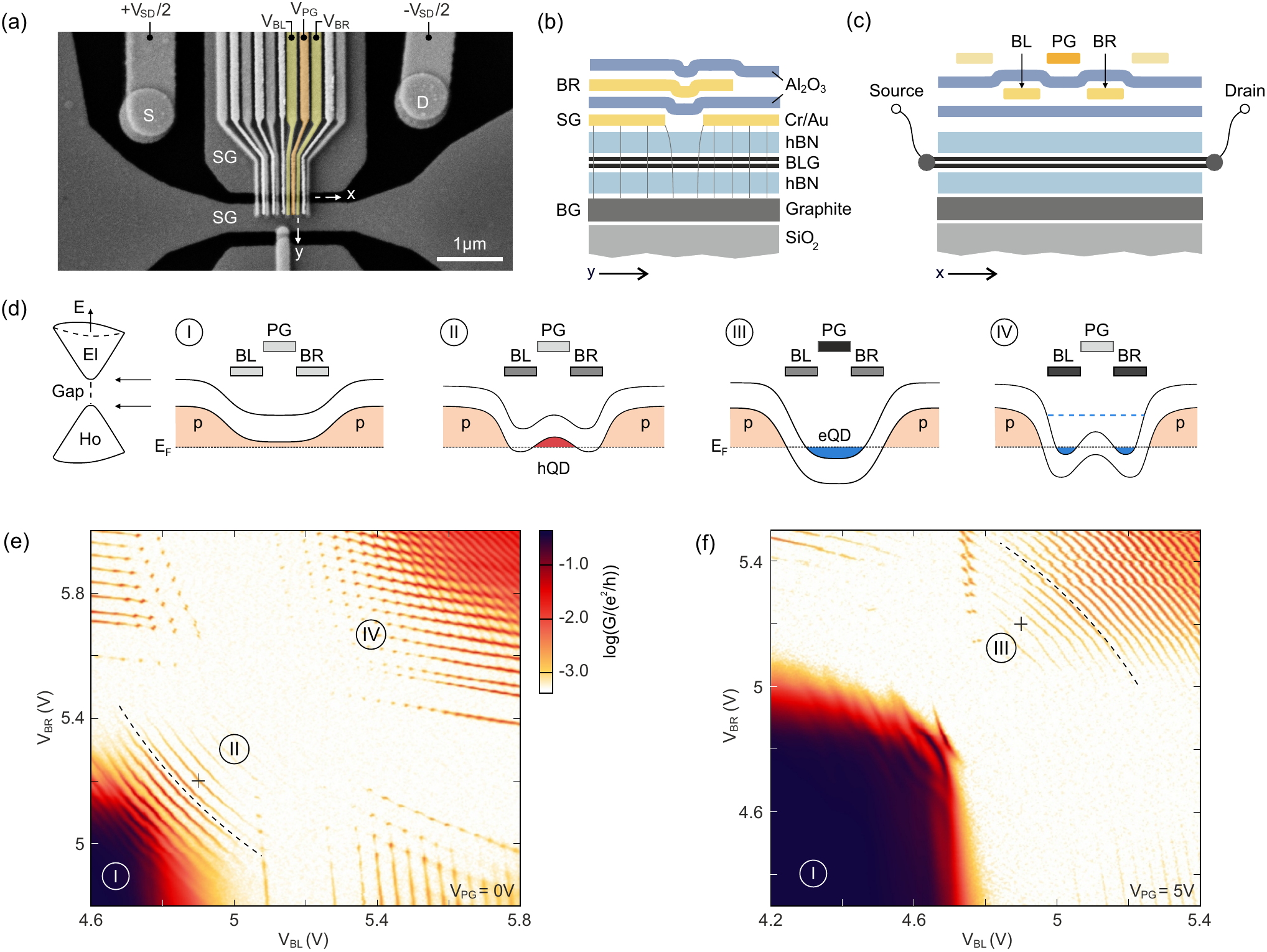}
\caption[Fig01]{
\textbf{(a)} {Scanning electron microscopy image showing the geometry of the QD device. The split gates (SGs) define the conducting channel, which can be modulated by voltages on the finger gates (highlighted by false colors). The ohmic source and drain contact to the BLG are labelled by S and D, respectively.
\textbf{(b)} and \textbf{(c)} Schematic cross-sections through the device along different directions (see white arrows in panel (a)). The hBN/BLG/hBN heterostructure rests on a graphite flake, which serves as a back gate (BG) and is placed on a SiO$_2$/Si substrate.
The gate stack placed on top of the heterostructure consists of three metal (Cr/Au) layers, insulated from each other by Al$_2$O$_3$-layers as gate dielectric: The SGs are followed by two layers of finger gates, some labeled here as barrier gates (BL, BR; yellow) and plunger gate (PG; orange). 
\textbf{(d)} Schematics of the valance and conduction band edge profiles along the p-doped channel illustrating the different regimes (I, II, III and IV) set by the barrier gate and plunger gate voltages (the darker grey the finger gate, the higher the applied voltage).
The band edges separate electron (El) and hole (Ho) states.
\textbf{(e)} Charge stability diagram showing the conductance as a function of the barrier gate voltages $V_\mathrm{BL}$ and $V_\mathrm{BR}$ at $V_\mathrm{PG} = 0$~V and $V_\mathrm{SD} = 200 \, \mathrm{\mu V}$.  
\textbf{(f)} Similar to panel (e) but at $V_\mathrm{PG} = 5$~V. The labels in (e) and (f) correspond to the ones in panel~(d), the dashed lines are described in the text and the black crosses mark the exact same barrier gate voltages.}
}

\label{f1}
\end{figure*}

Here, we report on the observation of the electron-hole crossover in a gate-defined bilayer graphene QD, by independently controlling the tunneling barriers and the dot occupation. 
Two dedicated finger gates induce two tunneling barriers in a split-gate defined channel and define a single QD between the barriers. 
The electrochemical potential of the QD can then be independently controlled by a third finger gate. Thanks to the small band gap in BLG, moderate gate voltages allow the tuning of the QD from the few-hole regime through the band gap into the few-electron regime. Finite-bias spectroscopy measurements allow to determine the energy scales of the system such as the charging energies (3-5~meV) and the band gap (around $25$~meV), which is in good agreement with the applied displacement field. 
Finally, we show that when applying an out-of-plane magnetic field electrons and holes are polarized in different valleys (K$^+$ and K$^-$) due to a large valley $g$-factor but an opposite sign of the topological orbital magnetic moment. The experimentally extracted valley $g$-factor ($g_v \approx 17$) is in good agreement with tight-binding calculations.

The geometry of our device is shown in Figs.~1a-c. A dry van der Waals stacking technique~\cite{Engels2014Sep,Wang2013Nov} is used to encapsulate a BLG flake by two hBN flakes (each $\approx 25$~nm thick). This stack is then placed on a graphite flake serving as back gate and ohmic source/drain contacts are made by etching holes into the stack (see Fig.~1a).
Next metallic (Cr/Au) split gates (SGs) with a lateral separation of $\approx 130$~nm are deposited (cross-section shown in Fig.~1b), which define the 2~$\mu$m long channel where the QDs will be formed (see top view in Fig.~1a and cross-section along the channel in Fig.~1c).
This structure is covered by a $25$~nm thick layer of atomic layer deposited (ALD) Al$_2$O$_3$. On top we place a first set of finger gates with a width of 70~nm and a pitch of $150$~nm, which later serve as left and right barrier (BL and BR) gates (see cross-section in Figs. 1b,c). 
Finally we deposit a second ALD Al$_2$O$_3$ layer and place a second set of interdigitated finger gates (again with a width of 70~nm. One of these gates will later be used as a plunger gate (PG). The different metall layers (all Cr/Au) have a thickness of $\approx 25$~nm. 
All measurements are performed in a ${}^3\mathrm{He}/{}^4\mathrm{He}$ dilution refrigerator at a base temperature of 10~mK, using \new{standard DC and low frequency lock-in (MFLI Zurich Inst.) measurement techniques.}

\begin{figure*}[!thb]
\centering
\includegraphics[draft=false,keepaspectratio=true,clip,width=1\linewidth]{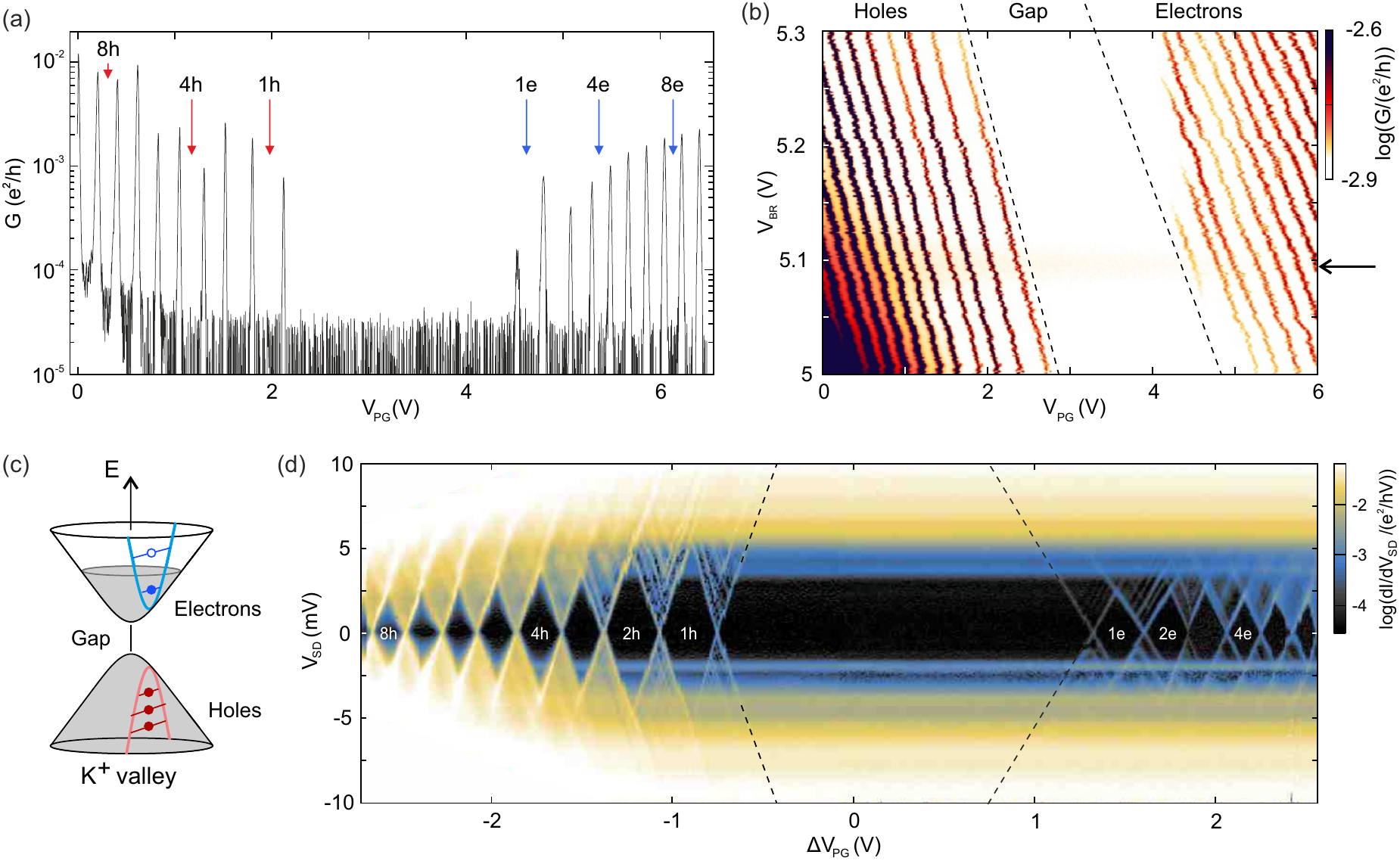}
\caption[Fig02]{
\textbf{(a)} Coulomb resonances as function of the plunger gate voltage $V_\mathrm{PG}$ and fixed $V_\mathrm{BR} = 5.09$~V and $V_\mathrm{BL} = 4.9$~V ($V_\mathrm{SD} = 200~\mu$V). The red and blue arrows (incl. labels) mark the occupations of hole and electron dot states separated by the band gap.
\textbf{(b)} Charge stability diagram showing the device conductance as a function of $V_\mathrm{PG}$ and $V_\mathrm{BR}$ for $V_\mathrm{BL} = 4.9$~V, highlighting the hole, gap and electron regime.
\textbf{(c)} Schematic band structure around the K$^+$ valley, highlighting possible QD states of an electron and hole QD. The gap originates from the transverse displacement field. The Fermi level lies in the conduction band such that the QD is filled with one electron.
\textbf{(d)} Finite bias spectroscopy measurement (along the arrow in panel (b)) showing the electron-hole crossover and the first Coulomb diamonds (see labels for occupation) of the few-electron and hole QD. 
For highlighting the charge neutrality point (set by the center of the band gap diamond) we plot this data as function of the relative gate voltage change $\Delta V_\mathrm{PG}$ with respect to the charge neutrality point ($\Delta V_\mathrm{PG}=0$~V).
}
\label{f2}
\end{figure*}

To confine %force 
the charge carriers to the 130~nm wide channel connecting the source and drain leads, we open a band gap in the BLG areas below the 
split gates. This is achieved by applying a transverse displacement field of 0.35~V/nm, which we induce by a back gate voltage of $V_\mathrm{BG} = -2 \, \mathrm{V}$ and a split gate voltage of $V_\mathrm{SG} = 2.15 \, \mathrm{V}$.  
The displacement field breaks the inversion symmetry of the BLG resulting 
%Following Ref. [] this results
in a band gap of around 28~meV ~\cite{McCann2006,Slizovskiy2019Dec} (see left illustration in Fig. 1d,I). Moreover, the similar magnitudes of the applied voltages make sure that below the SGs the Fermi level is within the band gap.
The negative BG voltage leads to a p-doping of the channel as there is no SG above the channel for compensating the field-effect induced charge carrier density.
This task, however, can now be performed by the different finger gates, which allow full electrostatic control of the shape of the band profile along the channel as depicted in Fig.~1d.
For example, while keeping the PG voltage constant ($V_\mathrm{PG}=0~$V) we can use the left and right barrier gate, BL and BR, to pinch-off the current through the channel. In Fig.~1e we show a corresponding charge stability diagram, i.e. the conductance (in log-scale) as function of positive BL and BR voltages ($V_\mathrm{BL}$ and $V_\mathrm{BR}$), showing a number of different regimes (see labels). In the lower left corner (regime I) the conductance is on the order of $e^2/h$ and the Fermi level is throughout the entire channel below the valence band edge (see Fig. 1d,I). 
When increasing the voltages on the left and right barrier gate we locally push the band edges down in energy such that below BL and BR the Fermi level is within the band gap (see Fig. 1d,II). This forms a hole QD (hQD) coupled by two tunneling barriers to the p-doped source and drain reservoirs. 
In the charge stability diagram a strongly reduced conductance and nearly periodic Coulomb resonances with a slope of around -1 (regime II) appear, meaning that both barrier gates are tuning the QD states equally well. 
The slightly negative curvature of the Coulomb resonances in the $V_\mathrm{BL}$-$V_\mathrm{BR}$-plane (see dashed line in Fig.~1e) are typical for gate-defined hole QDs: With increasing voltage at one barrier gate the QD is shifted towards the other barrier gate, which then couples more strongly, resulting in an overall asymmetric gate coupling.

\begin{figure*}[!thb]
\centering
\includegraphics[draft=false,keepaspectratio=true,clip,width=0.95\linewidth]{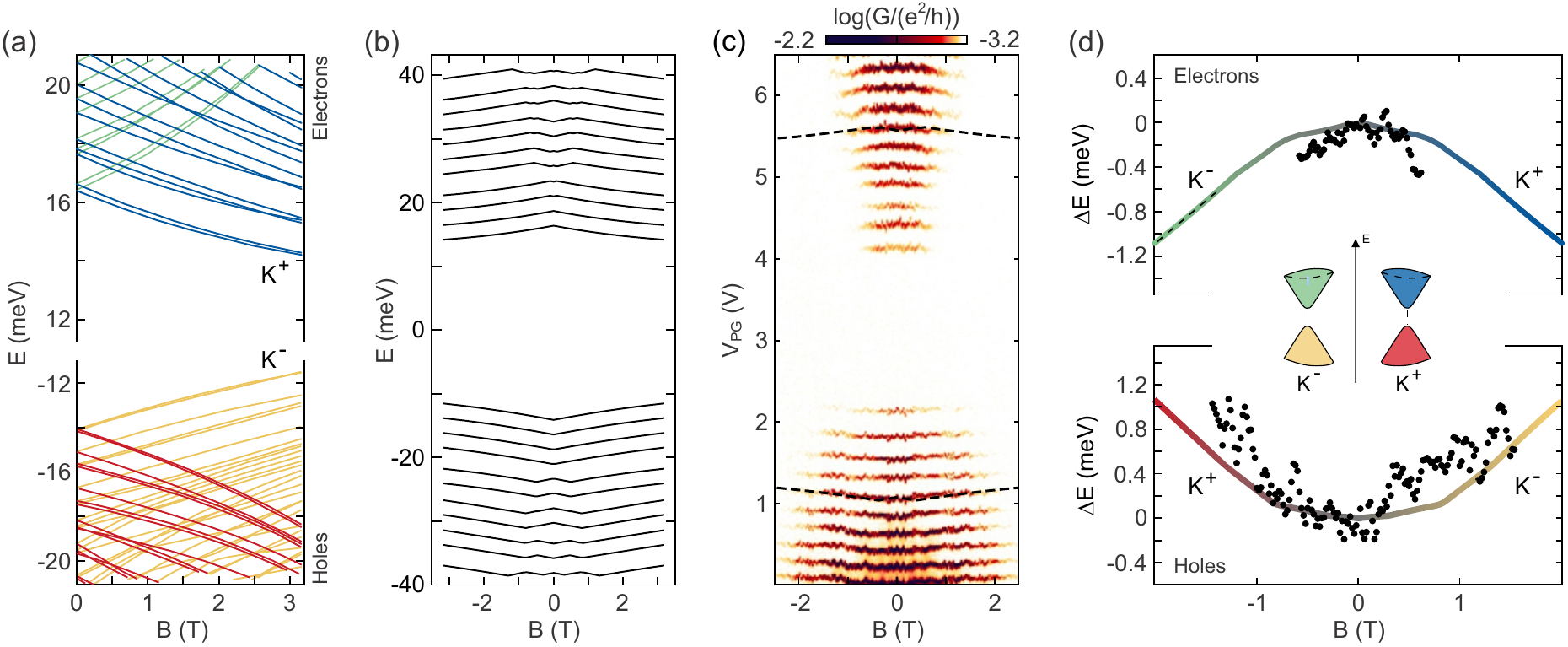}
\caption[Fig03]{
\textbf{(a)} Calculated single particle spectrum of a BLG QD with a size of $100 \times 250$~nm as function of an out-of-plane magnetic field (see text). With increasing field, the electron and hole states become K$^+$ and K$^-$ polarized, respectively. \new{The valleys are assigned by the sign of the orbital magnetic moment (slope).}  At zero magnetic field, orbital degeneracies are observed (multiplets of colored lines at $B=0$~T).
\textbf{(b)} Coulomb peaks reproduced from (a) by adding a charging energy of $3.4$~meV to each single particle eigenstate in order to model the experimental  transport data. The two-fold spin and two-fold orbital degeneracy \new{(due to the choosen potential)} lead to a four-fold degenerate spectrum. 
\textbf{(c)} Measured Coulomb resonances in the electron and hole regime as a function of $V_\mathrm{PG}$ and an out-of-plane magnetic field, $B$ ($V_\mathrm{SD}=400~\mu$V). \new{The dashed line is a guide to the eye and the positive/negative slope of holes/electrons as function of positive $B$-fields reflects the opposite net valley polarization.} 
\textbf{(d)} Energy shift of the Coulomb resonances (i.e. peaks) in panel (c) with respect to the peak position at $B = 0 \, \mathrm T$, averaged over the first ten charge transitions for the electron (top) and hole (bottom) regime.
The colored lines show the average energy shift of the first ten states obtained from the tight-binding calculation presented in panel (b). The colors of the lines indicate the increasing valley polarization with the color code indicated in the inset.
}
\label{f3}
\end{figure*}

By keeping the barrier gate voltages fixed (see cross in Figs. 1e and 1f) but changing the PG we can tune the system from a hole QD to an electron QD (eQD) configuration (Fig.~1d,III).
A corresponding charge stability diagram for $V_\mathrm{PG}=5~$V is shown in Fig.~1f. 
Again we see Coulomb resonances with a slope of about -1 (regime III around the cross in Fig.~1f).
The spacings of these Coulomb resonances are smaller compared to those of the hQD, which is due to the fact that the eQD is larger than the hQD (compare Fig.~1d,II with Fig.~1d,III). In addition, a slight positive curvature of the Coulomb resonances is observed in Fig.~1f, which fits well with an eQD: By increasing the voltage at one of the barrier gates, the band edges are pushed down locally and the QD shifts exactly in the direction of this gate, which leads to an enhanced gate coupling and an increasing asymmetry in the coupling of BL and BR.
Note that even in the case of $V_\mathrm{PG}=0~$V at very high barrier gate voltages ($V_\mathrm{BL}, V_\mathrm{BR} \gtrsim 5.8$~V) an even larger eQD can be formed (see Fig.~1e and blue dashed line in Fig.~1d,IV). 
However, when using the barrier gates to tune from this configuration towards the hQD, one passes through a double quantum dot regime consisting of two eQDs (see Fig.~1d,IV and regime IV in Fig.~1e), highlighting the importance of having the plunger gate for observing the electron-hole crossover in a single QD. Additional charge-stability diagrams for different plunger gate voltages are shown in the supplementary material (Fig.~S1).

The crossover from a hole to an electron QD becomes visible at best in the  conductance as a function of $V_\mathrm{PG}$ at constant barrier gate voltages. Such a measurement is shown in Fig.~2a, where the Coulomb peaks of the first 10 holes and the first 10 electrons as well as the BLG band gap can be identified (see labels). 
To prove the single quantum dot nature of these resonances we measure in Fig.~2b a charge-stability diagram as function of $V_\mathrm{PG}$ and $V_\mathrm{BR}$, highlighting two sets of Coulomb resonances with a linear slope (see dashed lines) separated by a gap. This gap separates the hQD from the eQD (see schematic in Fig.~2c) and is directly connected to the displacement-field induced band gap in BLG.   
The observed slopes of the Coulomb resonances provide the relative lever arm, i.e. the relative gate coupling of the plunger gate with respect to the barrier gate BR.
For instance, the plunger gate couples to the hQD about $3.5$ times worse than the BR gate (slope of around $-0.28$), which can mainly be explained by the larger vertical distance of the PG (see Fig.~1c) and partial screening by the barrier gates. 
This effect even gets enhanced when tuning to the eQD, where the BR gate couples around $5.5$ more strongly than the PG (slope of around $-0.18$), because the eQD is larger and extends more towards the barrier gates (see Fig.~1c,III).  

To extract energy scales of our system we performed finite bias measurements.
In Fig.~2d we show corresponding Coulomb-blockade
diamond measurements, i.e., measurements of the differential conductance ($G = $d$I/$d$V_\mathrm{SD}$) as a function of bias voltage $V_\mathrm{SD}$ and $V_\mathrm{PG}$ for fixed barrier gate voltages (taken in the same regime as the data presented in Fig.~2a). 
A large diamond associated with the band gap separates the hole QD from the electron QD, highlighting the few-hole and the few-electron regime.
Here, the first hole and the first electron exhibit lever arms of $\alpha^{1\mathrm{h}} = 0.016$~meV/mV and $\alpha^{1\mathrm{e}} = 0.009$~meV/mV. From the extension of the big diamond, i.e. the intersection of the corresponding charging lines (see dashed lines in Fig.~\ref{f2}d), a band gap of $\approx 23.5$~meV is extracted. This gap is in agreement with the one estimated from the applied transverse displacement field (see above).
%theory assuming an out-of-plane displacement field of $D = 0.28~$V/nm~\cite{McCann2006,Slizovskiy2019Dec}.
The band gap itself appears in this measurement as not completely blocking transport for larger source-drain voltage values. For $|V_\mathrm{SD}|$ values that exceed $\approx 2.5$~mV a background conductance sets in and increases step wise with $|V_\mathrm{SD}|$. This is not fully understood but the fact that this background conductance is independent of the plunger gate voltage suggests that it originates from a parallel channel formed under the split gates rather than from transport through the QD.

The Coulomb diamond measurements (Fig.~2d) also allow to determine the addition energies of the first hole, $E_\mathrm{add}^\mathrm{h}$ = 5.3~meV, and of the first electron, $E_\mathrm{add}^\mathrm{e}$ = 2.4~meV. With the help of the gate lever arms $\alpha^{1\mathrm{h}}$ and $\alpha^{1\mathrm{e}}$ we accordingly extract the QD capacitances of $C_\Sigma^\mathrm{h} = 31$~aF and $C_\Sigma^\mathrm{e} = 66$~aF. 
Using a simplified plate capacitor model consisting of a disk $25$~nm above the graphite back gate, we estimate an effective diameter of the hole and electron quantum dot of about $170$~nm and $250$~nm, respectively. %(see Supplementary material for details). 
This is a rather rough estimate, but the order of magnitude is reasonably consistent with the geometry of our device and reflects the different sizes of the hole and electron QD (compare Figs.~1d,II and~1d,III).

Next, we show that electrons and holes respond almost mirror-symmetrically to an out-of-plane magnetic field.
%
%Next, we show that electron and holes respond nearly symmetric to a perpendicular magnetic field.
%
%Next, we show that electron and holes respond differently to a perpendicular magnetic field. 
The Berry curvature~\cite{Berry} near the valley centers K$^{\pm}$ in gapped bilayer graphene induces a topological orbital magnetic moment~\cite{PhysRevLett.99.236809,PhysRevB.84.125427,PhysRevLett.106.156801} giving rise to an anomalously large effective valley $g$-factor. 
Just as the Berry curvature has different signs for the different valleys, as well as for electrons and holes, the sign of the topological magnetic moment changes accordingly. This leads to a different valley polarization of the confined electrons and holes when applying an out-of-plane magnetic field.

The valley-polarization can indeed be clearly seen when looking at the single particle spectrum of an electron/hole quantum dot as function of out-of-plane magnetic field (Fig.~3a). Here, we present tight-binding calculations of a BLG QD of experimental dimensions using a finite mass term~\cite{PhysRevB.80.165406}. Explicitly, we assume a soft box of dimensions $150\times 200$~nm, and additionally add a cosine and a small linear term, to account for a not perfectly flat confinement. Moreover, we add a displacement field to open a band gap of $\Delta = 20$~meV with on-site terms of $\pm \Delta/2$. 

At zero magnetic field each orbital state is fourfold degenerate due to the spin and valley degrees of freedom. 
Applying an out-of-plane magnetic field causes a valley splitting~\cite{PhysRevLett.99.236809}, where
states from one valley (see e.g. green lines in Fig.~3a) rapidly increase in absolute energy value (see green and red lines in Fig.~3a), while states from the other valley (blue and orange lines) evolve towards the lowest Landau levels. 
%, resulting in a valley polarization at low energy .
At higher $B$-fields, a net valley polarization is therefore established for the first QD states (see labels in Fig.~\ref{f3}a). This valley-polarization is different for electrons (K$^+$) and holes (K$^-$).

In order to compare theory with experiment we add to our model a phenomenological charging energy of $E_c = 3.4$~meV by shifting the $n^{\text{th}}$ eigenvalue of the single particle Hamiltonian by $n \cdot E_c$ (see Fig.~\ref{f3}b). 
The model shows not only the band gap but also the different sign of the topological orbital magnetic moment for electrons and holes, expressed by the different sign of the slope of the individual states. 
The steepness of the slope at high $B$-field is given by the valley $g$-factor. 
Our model can be directly compared to Coulomb resonance measurements recorded as function of plunger gate voltage and out-of-plane $B$-field (Fig.~3c).  
We observe that the Coulomb resonances in the hole and in the electron regime shift towards the gap with \new{opposite slopes} with increasing out-of-plane magnetic field, in good agreement with Fig.~3b. 
This shift originates from a net valley polarization of the QD due to the large valley $g$-factor. 
Although we cannot resolve individual level crossings in the experiment, which is mainly due to the rather large size of the QDs, we observe that for small magnetic fields there is a very low $B$-field dependence of the QD states. This fits well with the zig-zag lines in Fig.~3b, which are due to level crossings of states with different valleys (see Fig.~3a) showing up as small kinks in Fig.~3b. 
Note that with increasing magnetic field, tunneling through the QD is increasingly suppressed, resulting in lower Coulomb peak amplitudes.

For a quantitative comparison between experiment and theory and for extracting the valley $g$-factor $g_v$, we investigate the energy shift of the Coulomb peaks with respect to the peak position at $B = 0 \, \mathrm T$, averaged over the first ten holes and ten electrons, as a function of $B$-field, see Fig.~3d. The experimental data shows an increasing valley polarization, reflected in the increasing energy shift towards lower (higher) values for electron (hole) states, respectively.  %magnitude of the slope of the black lines in Fig.~\ref{f3}(d). 
The colored lines are the result of the  same averaging treatment applied to the tight-binding model calculation shown in Fig.~3b and are in good agreement with the experiment, highlighting the opposite valley polarization for electrons and holes (see labels and inset in Fig.~3d). The maximum extracted $g_v$ is $\approx 16$ for electrons (see dashed line in Fig.~3d) and $\approx 17$  for holes.

For achieving agreement between theory and experiment it is crucial that in the model system the QD potential is not 'flat' but exhibits a sizeable zero point energy of a few meV. This is because in a large and 'flat' QD the evolution of the lowest Landau level near the band gap would limit the slopes of the lowest individual states to $g_v \approx 3$~\cite{PhysRevB.93.085401}. This value is in contrast to the maximum extracted $g_v$ of around 17  which has also been observed in earlier works \cite{Kurzmann2019Jul}.
The slightly smaller average valley $g$-factor in the model compared to experiment might be due to the specific potential shape, strain~\cite{PhysRevB.101.085118}, additional 
valley splitting, or effects which are not captured by the single particle Hamiltonian.
The small electron hole asymmetry might be explained by the different QD size for electrons and holes~\cite{Knothe2020Jun}.

To conclude, we demonstrate the electron-hole crossover in a gate defined BLG QD by employing a gating scheme where three finger gates independently control the two tunnel barriers and the electrochemical potential of the QD. We tune the occupation of the QD from the few hole regime across the band gap into the few electron regime, where the extracted band gap of 23.5~meV is in good agreement with the gap induced by the transverse displacement field. Finally we show that electrons and holes have different signs of the topological orbital magnetic moments allowing for different valley polarization at moderate out-of-plane magnetic fields. This together with the large valley $g$-factor makes quantum systems based on combinations of coupled BLG electron and hole quantum dots interesting for valley and spin-valley qubits. \new{For future experiments, implementing a radio frequency charge detector would allow to study dynamical processes in BLG QDs and help to prove the occupation number of the QD, independently from the tunneling rates.} 
\newline
\newline
\textbf{Acknowledgements} 
This project has received funding from the European Union’s Horizon 2020 research and innovation programme under grant agreement No. 881603 (Graphene Flagship) and from the European Research Council (ERC) under grant agreement No. 820254, the Deutsche Forschungsgemeinschaft (DFG, German Research Foundation) under Germany’s Excellence Strategy - Cluster of Excellence Matter and
Light for Quantum Computing (ML4Q) EXC 2004/1
- 390534769, through DFG (STA 1146/11-1), and by
the Helmholtz Nano Facility~\cite{Albrecht2017May}. T.F. and F.L. acknowledge support by FWF project I-3827 and WWTF project MA14-002. The computational results were obtained using the Vienna Scientific Cluster (VSC). Growth of hexagonal boron nitride
crystals was supported by the Elemental Strategy Initiative conducted by the MEXT, Japan ,Grant Number
JPMXP0112101001, JSPS KAKENHI Grant Numbers
JP20H00354 and the CREST(JPMJCR15F3), JST.

\bibliography{literature}

\end{document}

%\preprint{APS/123-QED}

\author{L.~Banszerus}
\email{luca.banszerus@rwth-aachen.de.}
\affiliation{JARA-FIT and 2nd Institute of Physics, RWTH Aachen University, 52074 Aachen, Germany,~EU}%
\affiliation{Peter Gr\"unberg Institute  (PGI-9), Forschungszentrum J\"ulich, 52425 J\"ulich,~Germany,~EU}

\author{A.~Rothstein}
\affiliation{JARA-FIT and 2nd Institute of Physics, RWTH Aachen University, 52074 Aachen, Germany,~EU}%

\author{T.~Fabian}
\affiliation{Institute for Theoretical Physics, TU Wien, 1040 Vienna, Austria, EU}

\author{S.~M\"oller}
\author{E.~Icking}
\affiliation{JARA-FIT and 2nd Institute of Physics, RWTH Aachen University, 52074 Aachen, Germany,~EU}%
\affiliation{Peter Gr\"unberg Institute  (PGI-9), Forschungszentrum J\"ulich, 52425 J\"ulich,~Germany,~EU}

\author{S.~Trellenkamp}
\author{F.~Lentz}
\affiliation{Helmholtz Nano Facility, Forschungszentrum J\"ulich, 52425 J\"ulich,~Germany,~EU}

\author{D.~Neumaier}
\affiliation{AMO GmbH, Gesellschaft f\"ur Angewandte Mikro- und Optoelektronik, 52074 Aachen, Germany, EU}

\author{K.~Watanabe}
\author{T.~Taniguchi}
\affiliation{ 
National Institute for Materials Science, 1-1 Namiki, Tsukuba, 305-0044, Japan 
}%
\author{F.~Libisch}
\affiliation{Institute for Theoretical Physics, TU Wien, 1040 Vienna, Austria, EU}
\author{C.~Volk}
\author{C.~Stampfer}
\affiliation{JARA-FIT and 2nd Institute of Physics, RWTH Aachen University, 52074 Aachen, Germany,~EU}%
\affiliation{Peter Gr\"unberg Institute  (PGI-9), Forschungszentrum J\"ulich, 52425 J\"ulich,~Germany,~EU}%

%% CS: Possible alternative title?
\title{Supporting information - Electron-hole crossover in gate-controlled bilayer graphene quantum dots}%

\date{\today}% It is always \today, today,

%\begin{abstract} 
%\end{abstract}

\maketitle

\renewcommand{\thefigure}{S\arabic{figure}}
\setcounter{figure}{0}

\begin{figure*}[!thb]
\centering
\includegraphics[draft=false,keepaspectratio=true,clip,width=0.8\linewidth]{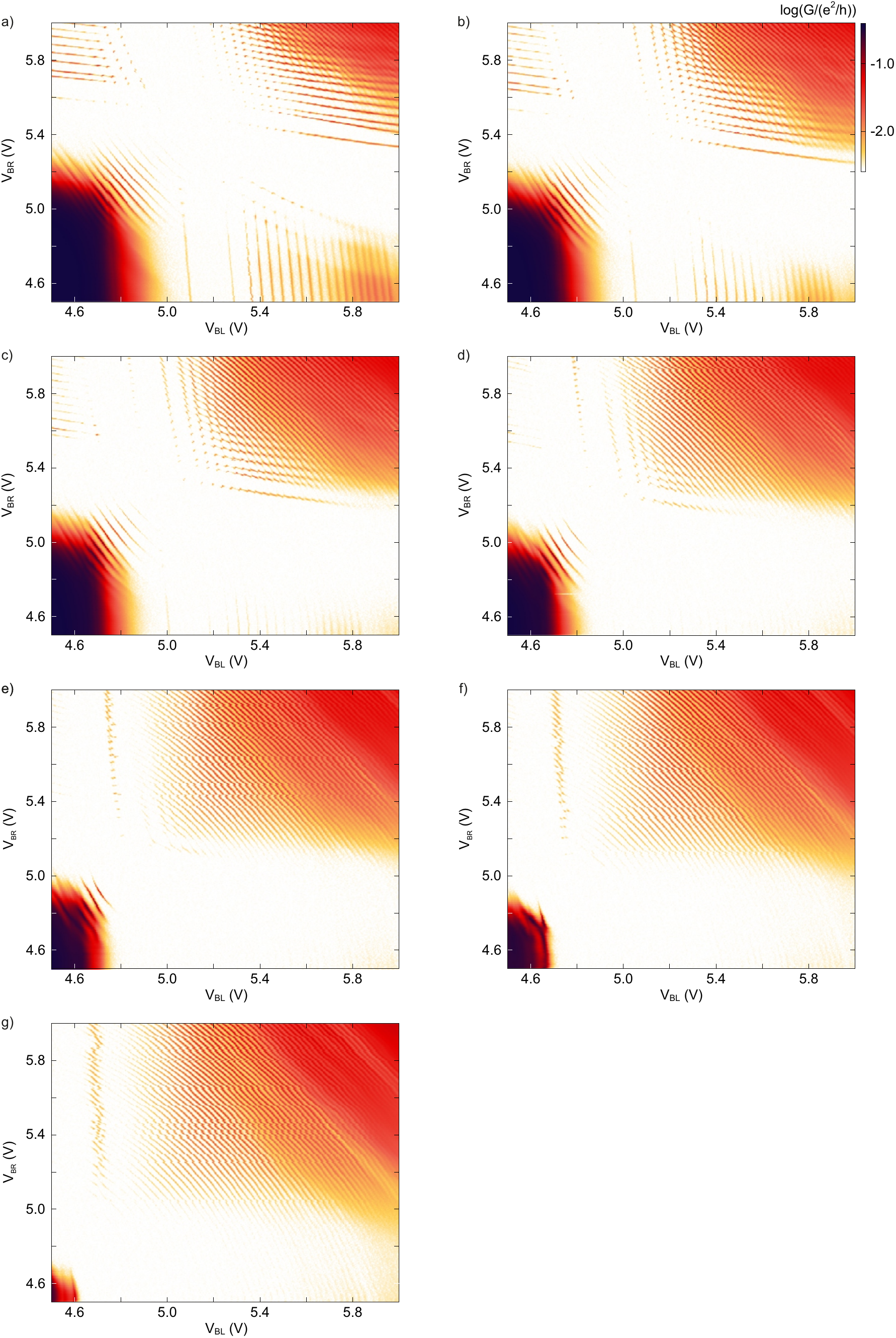}
\caption[Fig02]{Gallery of charge stability diagrams for different plunger gate voltages similar to Fig.1~e and f of the main text. The conductance is plotted as function of $V_\mathrm{BL}$ and $V_\mathrm{BR}$. With increasing $V_\mathrm{PG}$ a transition from a hole QD (lower left hand corner in panels a-d) to an electron QD (panels e-g) is observed.
\textbf{(a)} $V_\mathrm{PG} = 0~$V,
\textbf{(b)} $V_\mathrm{PG} = 1~$V,
\textbf{(c)} $V_\mathrm{PG} = 2~$V,
\textbf{(d)} $V_\mathrm{PG} = 3~$V,
\textbf{(e)} $V_\mathrm{PG} = 4~$V,
\textbf{(f)} $V_\mathrm{PG} = 5~$V and
\textbf{(g)} $V_\mathrm{PG} = 6~$V.
}
\label{fS1}
\end{figure*}

\begin{figure}[!thb]
\centering
\includegraphics[draft=false,keepaspectratio=true,clip,width=\linewidth]{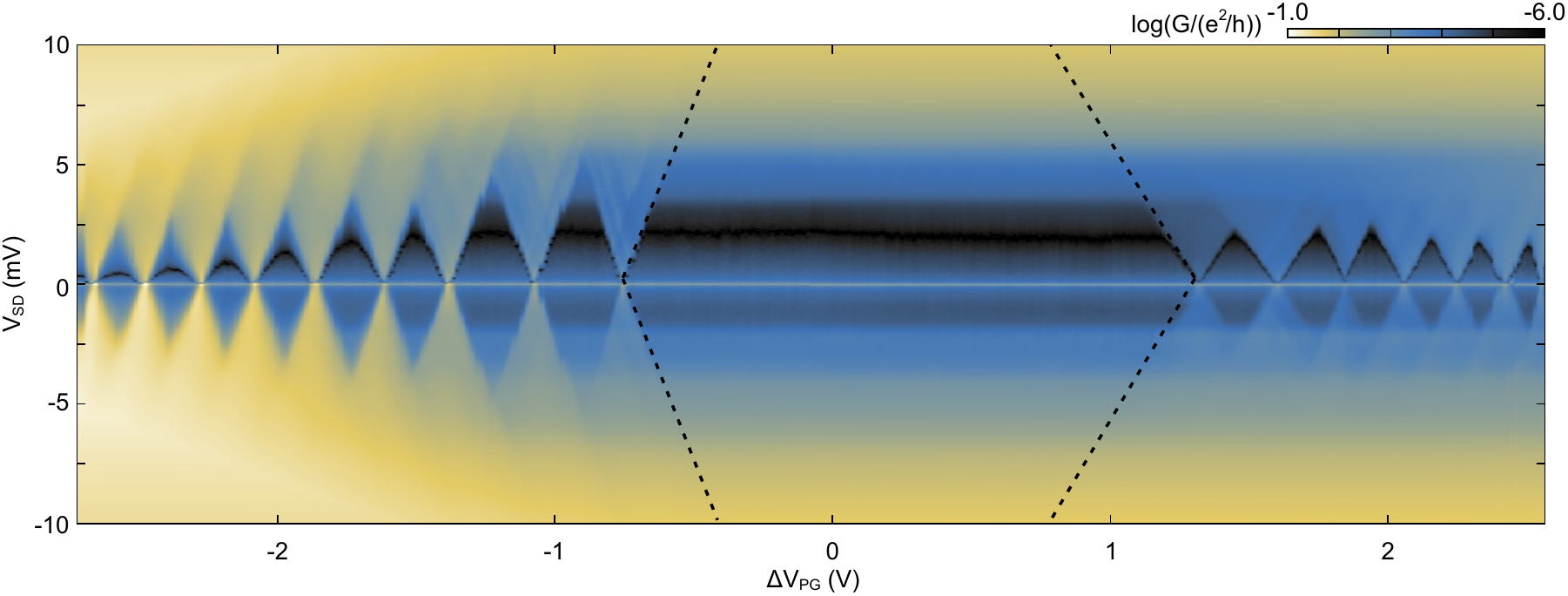}
\caption[Fig04]{Total logarithmic conductance as function of the applied bias and plunger gate voltage. The black dashed lines mark the charging of the first electron and hole. In between a band gap of $\approx23.5$~meV is observed.
}
\label{fS4}
\end{figure}

	\section{Theoretical Model}
	
	We set up a $240$~nm $\times$ $240$~nm large bilayer graphene flake in the parametrisation given in Ref.~\cite{PhysRevB.80.165406}.
	The magnetic field is implemented as Peierls phase~\cite{Peierls1933}.
	We compute the 50 eigenenergies closest to zero energy of this large, sparse Hamiltonian, which is best done with Krylov-subspace methods~\cite{Arpack}.
	Care has to be taken to avoid edge-states along the zig-zag edge, and therefore we use cylindrical boundary conditions.
	
	A finite mass term $m \; \sigma_z$, where $\sigma_z$ acts on sublattice, defines the quantum dot.
	Explicitly, we use for the mass 
	\begin{align}
		m (x,y) = & 0.5  \left( 1 + \exp \left( 90 - \sqrt{x^2}\right) /5  \right)^{-1} + 0.5  \left( 1 + \exp \left( 110 - \sqrt{y^2}\right) /5 \right)^{-1} \\
			& + .015 \left(  1 - \cos (3.5 \pi y / 120) \right) +  y / 120000,
	\end{align}
	where lengths $x,y$ are in nm and $m$ itself is in eV.
	The first two terms provide a smooth confinement potential.
	Already with these terms qualitative agreement with experiment is reached.
	The last terms are introduced to add a sizable zero-point energy to the system, which increases the valley $g$ factor of the lowest states, and provides several (in our case two) almost-degenerate minima, which leads to faster valley polarization, in good agreement with experiment.
	
		\section{Additional experimental data}

	\bibliography{literature}